\documentclass[preprints,article,accept,oneauthor,pdftex]{Definitions/mdpi} 
\firstpage{1} 
\pubvolume{xx}
\issuenum{1}
\articlenumber{5}
\pubyear{2019}
\copyrightyear{2019}
\history{Received: date; Accepted: date; Published: date}

\pdfoutput=1
\Title{L\'evy HBT results at NA61/SHINE}
\Author{Barnab\'as P\'orfy}
\address[1]{%
Wigner Research Centre for Physics, E\"otv\"os Lor\'and University, Budapest, Hungary; bporfy@cern.ch}

\abstract{Bose-Einstein (or HBT) momentum correlations reveal the space-time structure of the particle emitting
source created in high energy nucleus-nucleus collisions. In this paper we present the latest NA61/SHINE measurements of Bose-Einstein  correlations of identified pion pairs and their description based on L\'evy distributed sources in Be+Be collisions at 150\textit{A} GeV/\textit{c}. We investigate the transverse mass dependence of the L\'evy source parameters and discuss their possible interpretations.}


\begin{document}

\section{Introduction}
\begin{figure}
\includegraphics[width=1.0\linewidth]{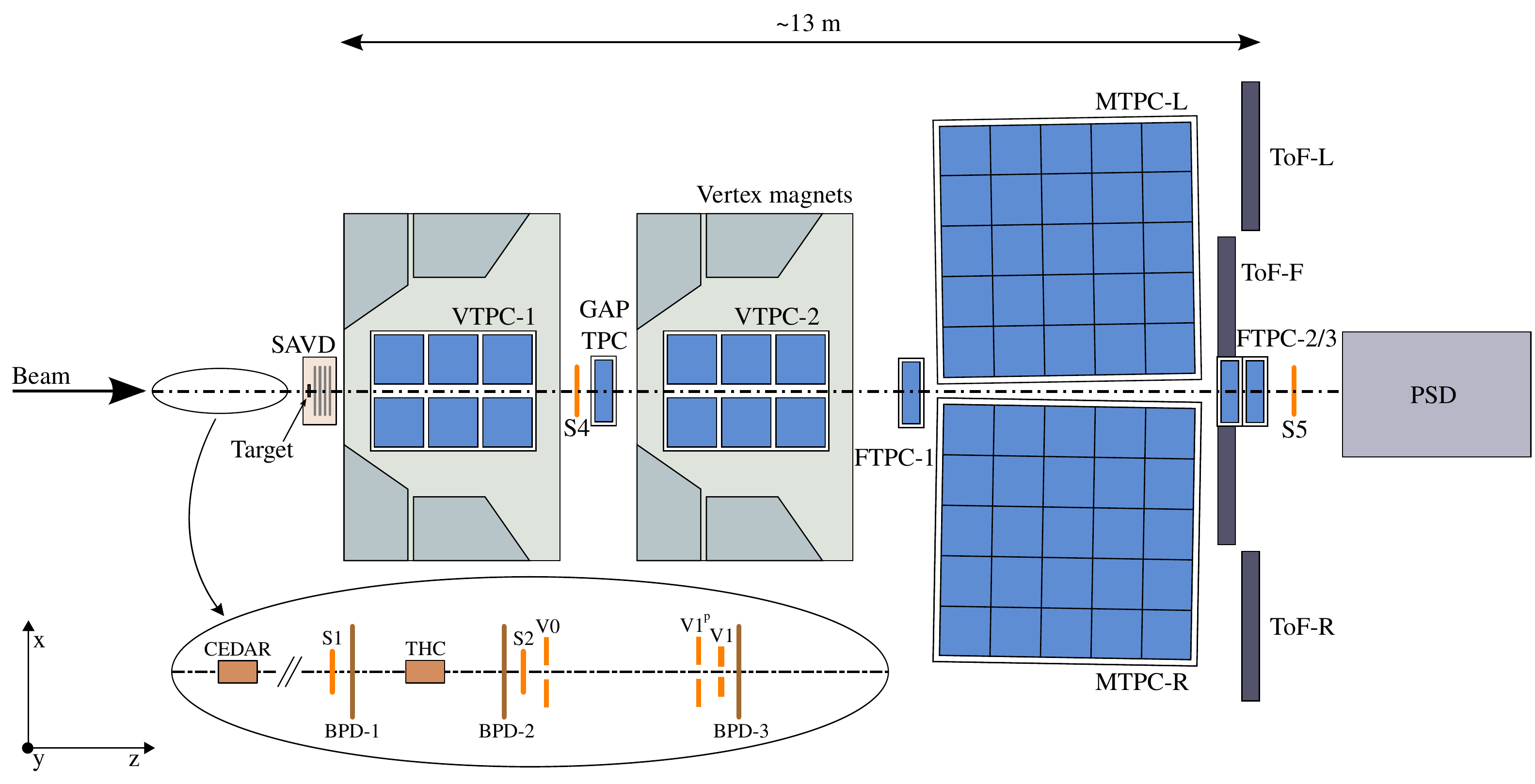}
\caption{The setup of the NA61/SHINE detector system.}
\label{f:na61etup}
\end{figure}

NA61/SHINE is a fixed target experiment at the CERN SPS. One if its main aims is to study the phase diagram of QCD. In order to accomplish that, different collision systems at multiple energies are investigated. The NA61/SHINE detector is equipped with 4 large Time Projection Chambers (TPC)~\cite{Abgrall:2014xwa}, these are covering the full forward hemisphere providing excellent tracking down to $p_T = 0$ GeV/\textit{c}. The experiment also features a modular calorimeter, located on the beam axis after the TPCs. This detector is called the Projectile Spectator Detector, it measures the forward energy which determines the collision centrality of the events. A setup of the NA61/SHINE detector system is shown in Fig.~\ref{f:na61etup}.

In order to study the QCD phase diagram and search for the Critical End Point (CEP), vastly different collision systems (p+p,p+Pb,Be+Be,Ar+Sc,Xe+La,Pb+Pb) are investigated at various beam momenta (13\textit{A}, 20\textit{A}, 30\textit{A}, 40\textit{A}, 75\textit{A} and 150\textit{A} GeV/\textit{c}). There are many observables to accomplish this goal. In the analysis described in this paper we measure Bose-Einstein (or HBT) correlations of identical pions in Be+Be collisions at 150\textit{A} GeV/\textit{c}. These, based on the principles of quantum-statistical correlations, reveal the femtometer scale structure of pion production, hence this field is often called femtoscopy.

\section{Femtoscopy, L\'evy sources and the Critical End Point}
The method of femtoscopy is based on the work of R. Hanbury Brown and R. Q. Twiss~\cite{HanburyBrown:1956bqd} as well as Goldhaber and collaborators~\cite{Goldhaber:1959mj}. The key relationship of this method shows that the spatial momentum correlations ($C(q)$) are related to the properties of the particle emitting source ($S(x)$, describing the probability density of particle creation) in the following way:
\begin{equation}
C(q) \cong 1 + | \tilde{S}(q) |^2,
\end{equation}
where $\tilde{S}(q)$ is the Fourier transform of $S(x)$, and $q$ is the momentum difference of the pair (dependence on the average momentum $K$ is suppressed here). See more details e.g. in Ref.~\cite{Adare:2017vig}. The usual assumption for the shape of the source is, based on the central limit theorem, Gaussian. A generalization of this assumption is to assume L\'evy distributed sources. A possible reason is that due to the expanding medium, the mean free path may increase and thus anomalous diffusion and L\'evy distributed sources may appear~\cite{METZLER20001,Csanad:2007fr}. Alternatively, due to critical fluctuations and the appearance of large scale spatial correlations, similar power-law tailed sources may be present~\cite{Csorgo:2005it}. Another reason for L\'evy distributed sources may be the fractal structure of QCD jets, as discussed in Ref.~\cite{Csorgo:2004sr}. Here we restrict our investigation to symmetric L\'evy distributions, as they have proven to provide a suitable description of Bose-Einstein correlations in nucleus-nucleus collisions~\cite{Adare:2017vig}. Furthermore, we restrict ourselves to describe the spatial part of the source, and the time dependence is absorbed through the connection of momentum difference $q$ and average momentum $K$ in case of identical particles:

\begin{align}
\vec{q}\vec{K}=q_0K_0.
\end{align}
Then the symmetric L\'evy distribution is characterized by two parameters: L\'evy scale parameter \textit{R} and the L\'evy exponent $\alpha$. The distribution is defined as follows:

\begin{equation}
\mathcal{L}(\alpha,R,r)=\frac{1}{(2\pi)^3} \int d^3q e^{iqr} e^{-\frac{1}{2}|qR|^{\alpha}},
\end{equation}
This distribution can be expressed analyticallly in two special cases. One is the already mentioned Gaussian distribution for $\alpha = 2$; furthermore, $\alpha=1$ leads to a Cauchy distribution. An important difference between L\'evy distributions and Gaussians is the presence of a power-law tail in case of  $\alpha < 2$, i.e. for large distances ($r$), the following holds:

\begin{equation}\label{eq:levygausdiff}
\mathcal{L}(\alpha,R,r) \sim r^{-(d-2+\alpha)},
\end{equation}
where \textit{d} represents the number of spatial dimensions. With L\'evy sources, the Bose-Einstein or HBT correlation functions can be expressed in the following way:

\begin{equation}
C(q) = 1 + \lambda \cdot e^{-(qR)^\alpha},\label{e:levyC0}
\end{equation}
where the $\lambda$ intercept parameter was introduced, which is defined as

\begin{equation}
\lambda = \lim_{q\rightarrow 0} C(q)
\end{equation}
where the $q\rightarrow 0$ extrapolation is done in experimentally available $q$ regions (limited from below by the two-track resolution in the given measurement). The Core-Halo model~\cite{Csorgo:1999sj,Csorgo:1995bi} may be utilized to understand this $\lambda$ intercept parameter. The Core-Halo model splits the source into two pieces. The core part contains the primordially created pions (directly from hadronic freeze-out) or from short lived (strongly decaying) resonances. The halo consists of pions created from longer lived (compared to the usual source size of a few femtometers) resonances and general background. In this picture, the $\lambda$ parameter turns out to be connected to the ratio of the Core and the Halo as follows:

\begin{equation}\label{e:corr_str}
\lambda = \left(\frac{N_{\rm{core}}}{N_{\rm{core}}+N_{\rm{halo}}}\right)^2
\end{equation}

Finally, let us come back to the above mentioned point of connecting L\'evy sources to the search for the Critical End Point of QCD. Critical points are characterized by critical exponents, one of which is the exponent of spatial correlations.  This appears because due to the second order phase transition at the CEP, the spatial correlation functions becomes a power-law with an exponent of $-(d-2+\eta)$ (where \textit{d} is the dimension and $\eta$ is the critical exponent of spatial correlations). We can see that the L\'evy exponent $\alpha$, given in Eq. (\ref{eq:levygausdiff}), defines a similar power-law, and hence $\alpha$ may be regarded as identical to the critical exponent $\eta$. See further discussions in Ref.~\cite{Csorgo:2005it}. Critical exponents are universal in the sense that they take the same values in case of physical systems belonging to the same universality class. It has been shown~\cite{Stephanov:1998dy} that the universality class of QCD is that of the 3D Ising model. The value of the critical exponent $\eta$ has been calculated to be $0.03631(3)$~\cite{El-Showk2014}. Alternatively, one may rely on the universality class of the 3D Ising model with a random external field, in which case an $\eta$ value of $0.5 \pm 0.05$ was calculated~\cite{Rieger:PhysRevB.52.6659}. Considering the previous statements, if we ``scan'' the phase diagram with different energies and systems and measure the values of the $\alpha$ exponent, we might be able to gain more information on the location and characteristics of the CEP.

\section{Measurement details}
In this measurement we analyzed the 0-20\% most central Be+Be collisions at 150\textit{A} GeV/\textit{c}.  This dataset consists of about 3 million events, which after various event and track quality selections was reduced to around 300 000 events. The track acceptance in this analysis was as follows. The rapidity region of analyzed particles is $0.85< \eta <4.85$ (corresponding to $|\eta|<2$ in the center-of-mass frame), the azimuthal coverage is $2\pi$; n this track sample, we identified pions based on their deposited energy $dE/dx$ in the TPC gas and charge obtained from the curvature of their trajectories in the magnetic field. We then analyzed negative pion pairs and positive pion pairs, as well as the combination of these two (i.e. created a dataset of identically charged pion pairs). These pairs were sorted into four $K_T$ (average pair transverse momentum) bins in the range of 0-600 MeV/\textit{c}. In each momentum bin, we measured the pair distribution of pairs from the same event, let us call this the $A(q)$ actual pair distribution. This contains quantumstatistical correlations, as well as many other residual effects related to kinematics and acceptance. To remove this undesirable effects, we created a mixed event for each actual event, by randomly selecting particles from other events of similar parameters, and making sure particles are each selected from a different event. Let us call the pair distribution from this sample $B(q)$, the background distribution. Then the correlation function is calculated as $C(q)=A(q)/B(q)$, provided a proper normalization is done in a $q$ range where quantumstatistical correlations are not expected. Let us mention here, that our analysis was done with a one-dimensional momentum difference variable $q$, calculated in the longitudinally co-moving system (LCMS), as in this frame, an approximately spherically symmetric source can be expected, furthermore, the extrapolation of $q\rightarrow 0$ is equivalent with the three dimensional case.

Final state effects are still present in the $C(q)$ correlation function. Among these, for pion pairs, the most important is the Coulomb effect, responsible for the repulsion of same charged pairs. It is usually handled by a so-called Coulomb correction as follows. The Coulomb correction for L\'evy type sources is a complicated numerical integral to calculate and fit, as discussed in Ref.~\cite{Adare:2017vig}. However it can be observed that the Coulomb correction does not strongly depend on the L\'evy exponent $\alpha$. Hence we can use then the approximate formula published by the CMS Collaboration in Ref.~\cite{Sirunyan:2017ies}, valid for $\alpha = 1$:

\begin{align}
K_{\textrm{Coulomb}}(q,R) &= \textrm{Gamow}(q)\cdot \left(1+\frac{\pi\eta(q) q\frac{R}{\hbar c}}{1.26+q \frac{R}{\hbar c}}\right),\textnormal{ where}\label{e:coulcorr}\\
\textrm{Gamow}(q) &= \frac{2\pi\eta(q)}{e^{2\pi\eta(q)-1}} \;\textnormal{ and }\;
\eta(q) = \alpha_{\textrm{QED}}\cdot\frac{\pi}{q},
\end{align}
where $\eta(q)$ is the so-called Sommerfeld parameter, and $\alpha_{\rm QED}$ is the fine-structure constant (neither should be confused with the above discussed exponents $\eta$ and $\alpha$). Utilizing the usual Bowler-Sinyukov method (i.e. Coulomb correcting only for the Core part of the source) as indicated in Refs.~\cite{Sinyukov:1998fc,Bowler:1991vx}, one obtains:

\begin{equation}\label{e:fitfunc}
C(q) = N\cdot\left(1-\lambda + \lambda\cdot\left(1+e^{-(qR)^{\alpha}}\right)\cdot K_{\textrm{Coulomb}}(q)\right),
\end{equation}
with $N$ being a normalization parameter responsible for the proper normalization of the $A(q)/B(q)$ ratio. This is the final fit function we are using to describe our data.

As also found in Ref.~\cite{Adare:2017vig}, the L\'evy parameters ($\alpha, R, \lambda$) are highly correlated, and especially in a low statistics dataset, it is hard to determine them precisely. We might be able to reduce this correlation and the statistical uncertainty of the parameters, if we fix one of the three parameters to a well motivated value. The resulting statistical uncertainties and free parameter values are modified due to the additional physical assumptions used to fix one of the parameters. From a statistical point of view, a bootstrap type of method may also be used. However, our main aim with this is to see a more clear trend of the $m_T$ dependence of the parameters, with additional physical assumptions. One assumption is that $\alpha$ (i.e. the shape of the pion emitting source) is independent of $m_T$, with that we may fix $\alpha$ to a weighted average of the 4 $\alpha$ values obtained in free parameter fits performed in each $K_T$ bin. The other option is fixing \textit{R} with the following equation motivated by hydrodynamical predictions of the particle emission homogeneity length (essentially the HBT radii) in case of expanding fireballs~\cite{Csorgo:1995bi}. In this case, we fit the following equation to the $m_T$ dependence (where $m_T=\sqrt{m^2+K_T^2}$, i.e. the average transverse mass of the pair) of the $R$ L\'evy scale:

\begin{equation}\label{e:rfix}
R(m_T) = \frac{A}{\sqrt{1+m_T/B}}.
\end{equation}
Previous results with free parameter fits were shown in Ref.~\cite{Porfy:2019lru}, hence here we concentrate on the results of the above mentioned fixed parameter fits. We again note that our aim with fixing one of the parameters to a physically motivated value is to show the trend of the $m_T$ dependence of the parameters.

%%%%%%%%%%%%%%%%%%%%%%%%%%%%%%%%%%%%%%%%%%
\section{Results}

\begin{figure}[H]
\centering
\includegraphics[width=1\textwidth]{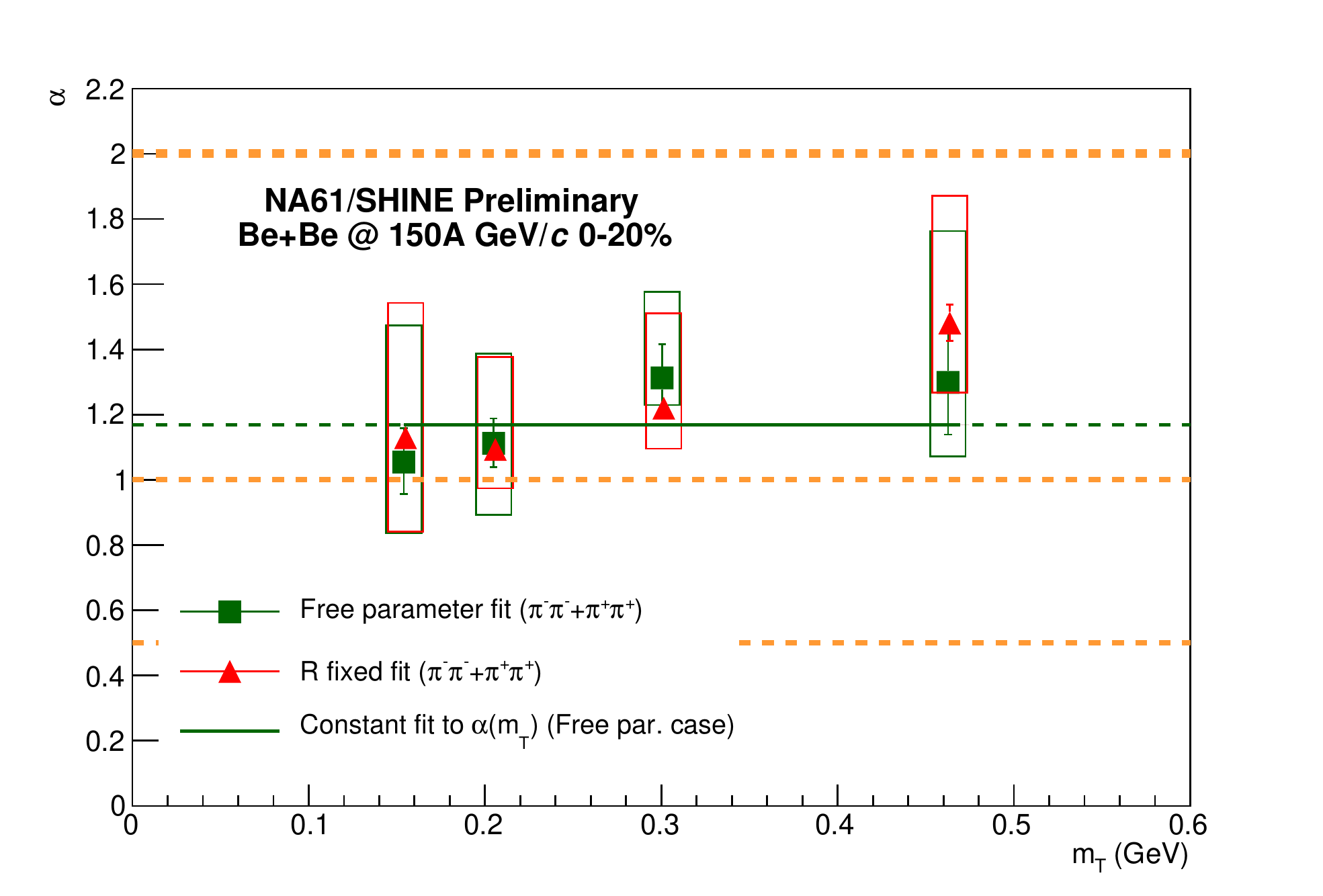}
\caption{L\'evy exponent $\alpha$ versus transverse mass: comparison between free parameter fit, fit with R fixed and fit with $\alpha$ fixed. The boxes represent systematic uncertainties. For each bin, the results are slightly shifted to the right for visibility, but they are in the same bin.}
\label{fig:ares}
\end{figure}

\begin{figure}[H]
\centering
\includegraphics[width=1\textwidth]{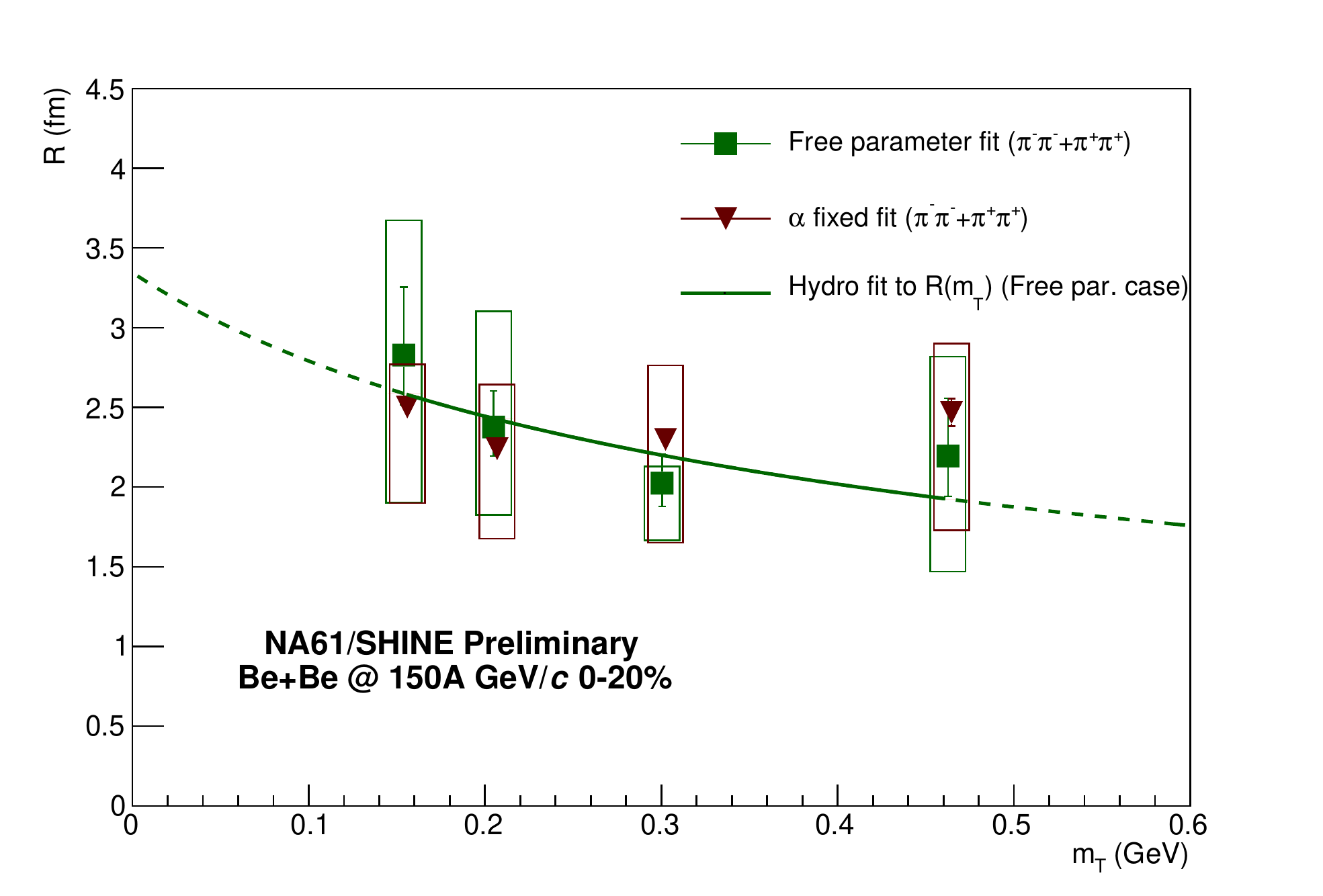}
\caption{Correlation strength $R$ versus transverse mass: comparison between free parameter fit, fit with $\alpha$ fixed and fit with R fixed. The boxes represent systematic uncertainties. For each bin, the results are slightly shifted to the right for visibility, but they are in the same bin.}
\label{fig:rres}
\end{figure}

\begin{figure}[H]
\centering
\includegraphics[width=1\textwidth]{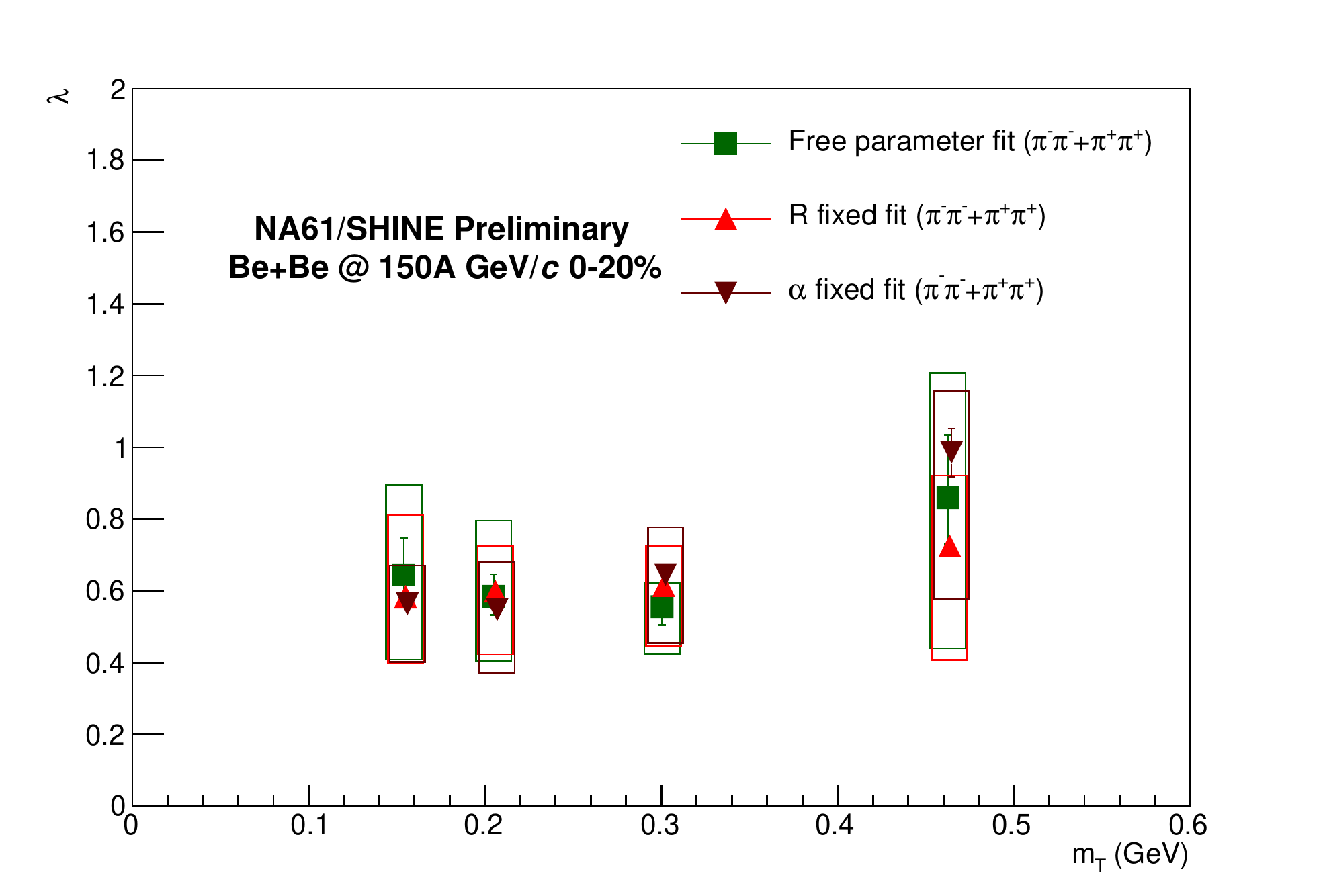}
\caption{L\'evy exponent $\lambda$ versus transverse mass: comparison between free parameter fit, fit with $\alpha$ fixed and fit with R fixed. The boxes represent systematic uncertainties. For each bin, the results are slightly shifted to the right for visibility, but they are in the same bin.}\label{fig:lres}
\end{figure}

First, the measured correlation functions were fitted with the above mentioned (Eq. \ref{e:fitfunc}) function with three free parameters ($\alpha, \lambda\;\rm{and}\; R$), as shown in  Ref.~\cite{Porfy:2019lru}. Using the results from the free parameter fit, we fitted a constant function to the $\alpha$ values for all $m_T$ bins, as well as the formula of Eq.~\eqref{e:rfix} to the $R$ values in each bin. Then we analyzed first with one parameter fixed. All results are shown in Figs.~\ref{fig:ares}-\ref{fig:lres}. Let us note here, that all the measurement settings (event selection, track selection, pair cuts, fitting interval) were varied systematically to obtain an estimate of systematic uncertainties. These, along with the statistical uncertainty of the parameters (obtained by the Minos algorithm) are shown  in Figs.~\ref{fig:ares}-\ref{fig:lres}.

The L\'evy stability exponent $\alpha$ determines the source shape, and a value of $2$ corresponds to a Gaussian source, a value of $1$ to a Cauchy source, and 0.5 is the conjectured value at the critical point. Our results, along with these special cases (dotted yellow lines), are shown in Fig.~\ref{fig:ares}. It is clear from the figure that the statistical uncertainties of $\alpha$ from the fixed $R$ fit are reduced by a factor 4-5; however, the values of $\alpha$ are similar in both cases. These values are far from the Gaussian case as well as the conjectured CEP value, motivating us to perform this measurement in different systems and at different energies as well.

The L\'evy scale $R$ determines the correlation length of pion pairs from the given system. From a simple hydrodynamical picture one obtains a $R \propto 1/\sqrt{m_T}$ type of affine linear dependence, as already mentioned above. This, i.e. Eq. (\ref{e:rfix}), describes the free parameter fit data points of Figure \ref{fig:rres} well. The fixed $\alpha$ fits are within uncertainties compatible with the free parameter results, however, they suggest a more or less constant trend of $R$ versus $m_T$ in the higher $m_T$ bins. This motivates us to perform the same measurement in collision systems with larger multiplicities (where statistical uncertainties are expected to be reduced proportionally to the square of the mean multiplicity).

The last parameter to study, is the correlation strength parameter $\lambda$, given in Eq.~(\ref{e:corr_str}). The transverse mass dependence of $\lambda$ is shown in Fig.\ref{fig:lres}. Comparing the three different fits (free parameter fit, fixed $\alpha$ fit, fixed $R$ fit), it is visible that $\lambda$ in free parameter fit is compatible (within statistical uncertainties) with both other cases. And all fits show a roughly constant $\lambda(m_T)$ trend. This is in contrast to the findings at RHIC, see e.g. the compilation in Ref.~\cite{Vertesi:2009wf}. This finding is however compatible with previous SPS measurements (in different systems), see e.g. Ref.~\cite{Beker:1994qv}.

%%%%%%%%%%%%%%%%%%%%%%%%%%%%%%%%%%%%%%%%%%
\section{Conclusions}

We reported above on the NA61/SHINE measurement of one-dimensional identified two-pion Bose-Einstein correlation functions in the 0-20\% most central Be+Be collisions at 150\textit{A} GeV/\textit{c}. We compared free parameter fits to fixed parameter fits, to reduce statistical uncertainty of the physical parameters ($\alpha, \lambda\;\rm{and}\;R$). We found that the results from the free parameter fits and the fixed parameter fits are similar, but the statistical uncertainty of each parameter is reduced by a large factor. The aim of this excercise was to show the trends of the parameters with fixing one parameter to a physically motivated value. Our results confirmed that in this collision system and at this collision energy, the L\'evy exponent $\alpha$ is far from the Gaussian case, as well as from the conjectured value at the Critical End Point. We furthermore found that the $R(m_T)$ dependence is compatible with hydro predictions, and $\lambda(m_T)$ may show different patters at RHIC and SPS energies. These findings will be subsequently investigated in other collision systems as well.

\funding{B. P\'orfy was supported by the NKFIH grants FK123842 and FK123959.}

%%%%%%%%%%%%%%%%%%%%%%%%%%%%%%%%%%%%%%%%%%
\acknowledgments{The author would like to thank the NA61/SHINE collaboration.}

%%%%%%%%%%%%%%%%%%%%%%%%%%%%%%%%%%%%%%%%%%
\conflictsofinterest{The author declares no conflict of interest.}

%%%%%%%%%%%%%%%%%%%%%%%%%%%%%%%%%%%%%%%%%%
%% optional
\abbreviations{The following abbreviations are used in this manuscript:\\

\noindent 
\begin{tabular}{@{}ll}
QCD & Quantum chromodynamics\\
CERN & European Organization for Nuclear Research\\
SPS & Super Proton Synchrotron\\
HBT & Hanbury Brown and Twiss\\
CEP & Critical End Point
\end{tabular}}

\reftitle{References}

%\bibliography{HBT_ref}

\end{document}